\documentclass[aip,apl,twocolumn,noshowpacs,superscriptaddress,amsmath, amssymb, reprint]{revtex4-1}
\usepackage{graphicx}
\usepackage{multirow}
\usepackage{color}
\usepackage{amsmath}

\newcommand{\GeTeSbTe}{GeTe/Sb$_2$Te$_3$}
\newcommand{\dash}{\nobreakdash-}

\begin{document}

\title{Electronic and Thermal Properties of \GeTeSbTe{} Superlattices by {\it{ab initio}} Approach: Impact of Van der Waals Gaps on Vertical Lattice Thermal Conductivity}

\author{Benoît~Sklénard}
\affiliation{Univ. Grenoble Alpes, CEA, Leti, F-38000, Grenoble, France}

\author{François~Triozon}
\affiliation{Univ. Grenoble Alpes, CEA, Leti, F-38000, Grenoble, France}

\author{Chiara~Sabbione}
\affiliation{Univ. Grenoble Alpes, CEA, Leti, F-38000, Grenoble, France}

\author{Lavinia~Nistor}
\affiliation{Applied Materials France, 38190 Bernin, France}

\author{Michel~Frei}
\affiliation{Applied Materials Inc., Santa Clara, CA 95054-3299, United States}

\author{Gabriele~Navarro}
\affiliation{Univ. Grenoble Alpes, CEA, Leti, F-38000, Grenoble, France}

\author{Jing~Li}
\email{Jing.Li@cea.fr}
\affiliation{Univ. Grenoble Alpes, CEA, Leti, F-38000, Grenoble, France}

\begin{abstract}
In the last decade, several works have focused on exploring the material and electrical properties of \GeTeSbTe{} superlattices (SLs) in particular because of some first device implementations demonstrating interesting performances such as fast switching speed, low energy consumption, and non\dash{}volatility. However, the switching mechanism in such SL\dash{}based devices remains under debate. In this work, we investigate the prototype \GeTeSbTe{} SLs, to analyze fundamentally their electronic and thermal properties by {\it{ab initio}} methods. We find that the resistive contrast is small among the different phases of \GeTeSbTe{} because of a small electronic gap (about $0.1$~eV) and a consequent semi\dash{}metallic\dash{}like behavior. At the same time the out\dash{}of\dash{}plane lattice thermal conductivity is rather small, while varying up to four times among the different phases, from $0.11$ to $0.45$~W~m$^{-1}$K$^{-1}$, intimately related to the number of Van der Waals (VdW) gaps in a unit block. Such findings confirm the importance of the thermal improvement achievable in \GeTeSbTe{} superlattices devices, highlighting the impact of the material stacking and the role of VdW gaps on the thermal engineering of the Phase\dash{}Change Memory cell.
\end{abstract}


\maketitle

Phase\dash{}Change Memory (PCM) is considered among the most promising non\dash{}volatile memory technologies. It has achieved in the last decade a high level of maturity, demonstrated by its commercialization for Storage Class Memory applications \cite{Cheng_2019} and its proven potential to become the mainstream solution for the embedded automotive market \cite{Cappelletti_2020}. 
In order to target ultra\dash{}low programming current in the next generation of the technological nodes, the programming current reduction in PCM has been the object of several works, in particular focusing on the engineering of the deposited material stack.
Indeed, the optimization of the thermal conductivity of the phase\dash{}change material, could lead to a significant reduction of the current needed to perform the crystalline\dash{}to\dash{}amorphous reversible transition in the PCM device during the programming operations \cite{Song_2020, Kusiak_2016}. 
Among the studies targeting the power efficiency improvement in PCM, the solution based on \GeTeSbTe{} crystalline superlattices (SLs) was proposed as featuring fast and low\dash{}power programming operations compared to the traditional bulk Ge$_{2}$Sb$_{2}$Te$_{5}$ (GST) \cite{Tominaga_2011}. 

\begin{figure}
 \includegraphics[width=\columnwidth]{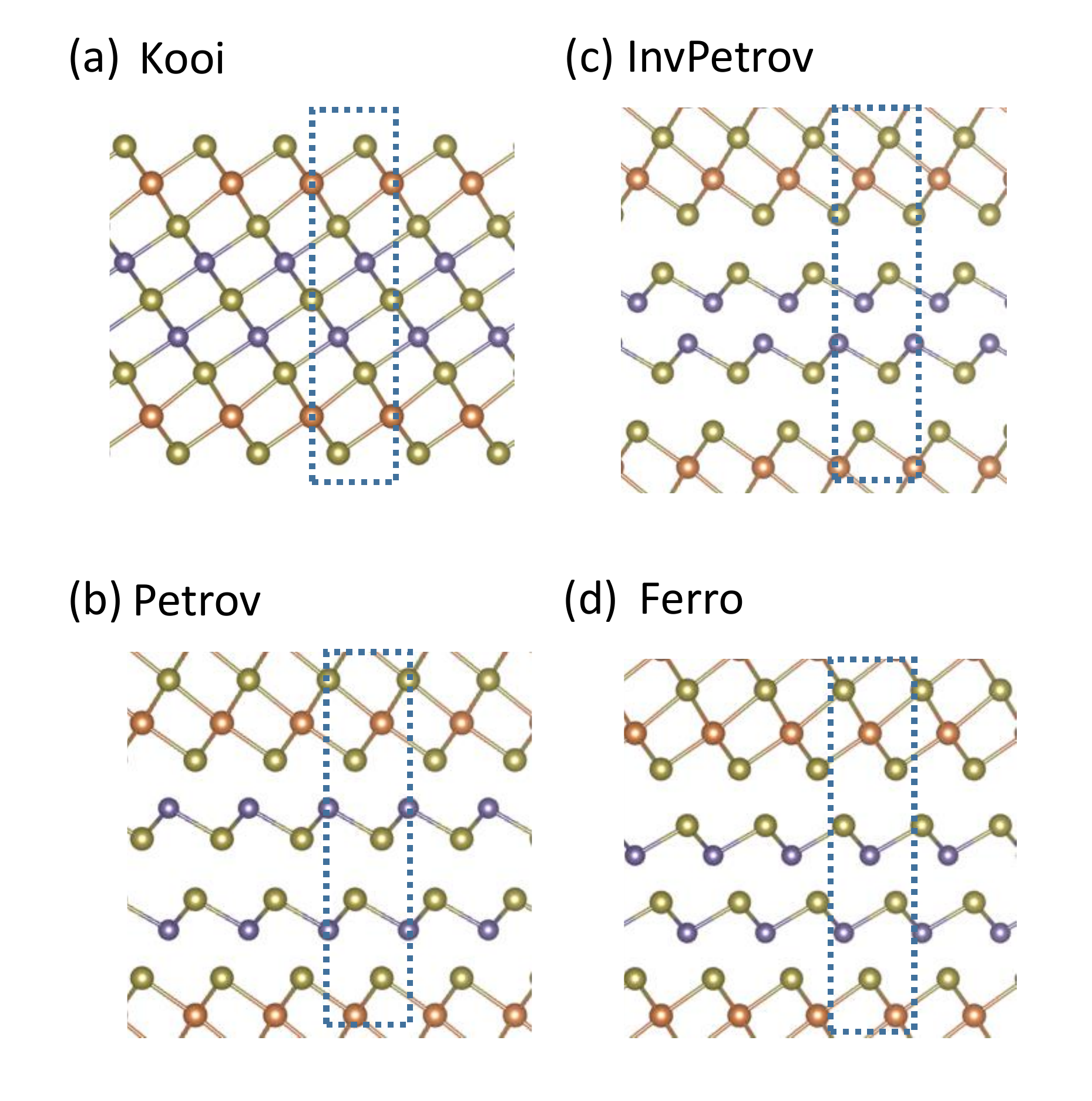}
 \caption{(a) GST Kooi structure and other hex GST phases: (b) Petrov, (c) InvPetrov, and (d) Ferro. The unit cells are indicated by a dashed blue rectangle. Ge, Sb, and Te atoms are in purple, red, and green respectively.
   In one unit cell, Kooi has single Te-Te VdW gap; Petrov and Ferro have one Te-Te VdW gap and two Ge-Te long bonds; and InvPetrov has two Te-Te VdW gaps and one Ge-Ge long bonds.}
 \label{str}
\end{figure}

Several different mechanisms have been evoked along the years to explain the transition between the SET state (i.e. low resistance state) and the RESET state (i.e. high resistance state) in such devices. 
In particular, a solid\dash{}to\dash{}solid transition without any melting of the active layer has been proposed. 
This first group of theories supports a crystal\dash{}to\dash{}crystal transition (i.e. hexagonal\dash{}to\dash{}hexagonal) based on the transition between two of the possible phases of hexagonal (hex) GST (i.e. between Petrov and Inverted\dash{}Petrov, or between Inverted\dash{}Petrov and Ferro, as reported in Fig.~\ref{str}), taking into account an atomic vertical diffusion (i.e. Ge diffusion) combined with a lateral motion of the sublayers \cite{Xiaoming_2015, Mitrofanov_2019}. 
More recently, in order to overcome the discrepancy with the experimental results and the oversimplification of previous theories, the reconfiguration of Van der Waals (VdW) gaps through the inversion of the SbTe planes between neighboring blocks has been proposed, justifying the presence of a local deviation in stoichiometry from quasi\dash{}binary compositions as observed in TEM experiments \cite{Kolobov_2017,Kowalczyk_2018}. 
In later reports, the motion of the stacking faults along the layer \cite{Chen2_2018} and the intermixing of Ge/Sb \cite{Saito_2019} have been proposed as mechanisms  behind the possible electronic properties change in \GeTeSbTe{} SLs. 

A second group of studies supports a thermal\dash{}based explanation for the reduced programming current in \GeTeSbTe{}. The phase\dash{}change transition is equivalent to standard bulk\dash{}based PCM, but is made more efficient thanks to the particular stack configuration of a SL. 
Preliminary works on amorphous SLs \cite{Chong_2006} have already evidenced an improved thermal efficiency in these layers, reinforced by several following works demonstrating the reduction of the thermal conductivity in SLs with respect to bulk layers \cite{Tong_2011, Long_2012}. 
Moreover, the anisotropy of the thermal conductivity that can be achieved in \GeTeSbTe{} SLs \cite{Campi_2017} could be extremely advantageous in reducing the power involved in the device programming operations \cite{Navarro_2018}. 
Recently, these studies have received even more support from the experimental demonstration in PCM devices of the higher thermal resistance of a SL with respect to a bulk device, attributed to the presence of multiple interfaces and multiple VdW gaps \cite{Boniardi_2019,Noe_2021,Kwon_2021,Khan_2021}.

In this letter, we report the investigation of the electronic and thermal properties of the different hex GST phases, including the most stable Kooi structure \cite{Kooi_2002} and the different \GeTeSbTe{} SLs, using {\it{ab initio}} methods. 
We highlight the small difference in electrical conductivity among the different phases and the presence of thermal conductivity anisotropy in all of them.
Finally, we show how the out\dash{}of\dash{}plane thermal conductivity is correlated with the total number of VdW gaps in the unit block, emphasizing importance of  stack engineering to control and enhance the presence of VdW gaps in the phase\dash{}change layer to achieve a more thermally efficient PCM device.

\begin{figure*}
 \includegraphics[width=1.8\columnwidth]{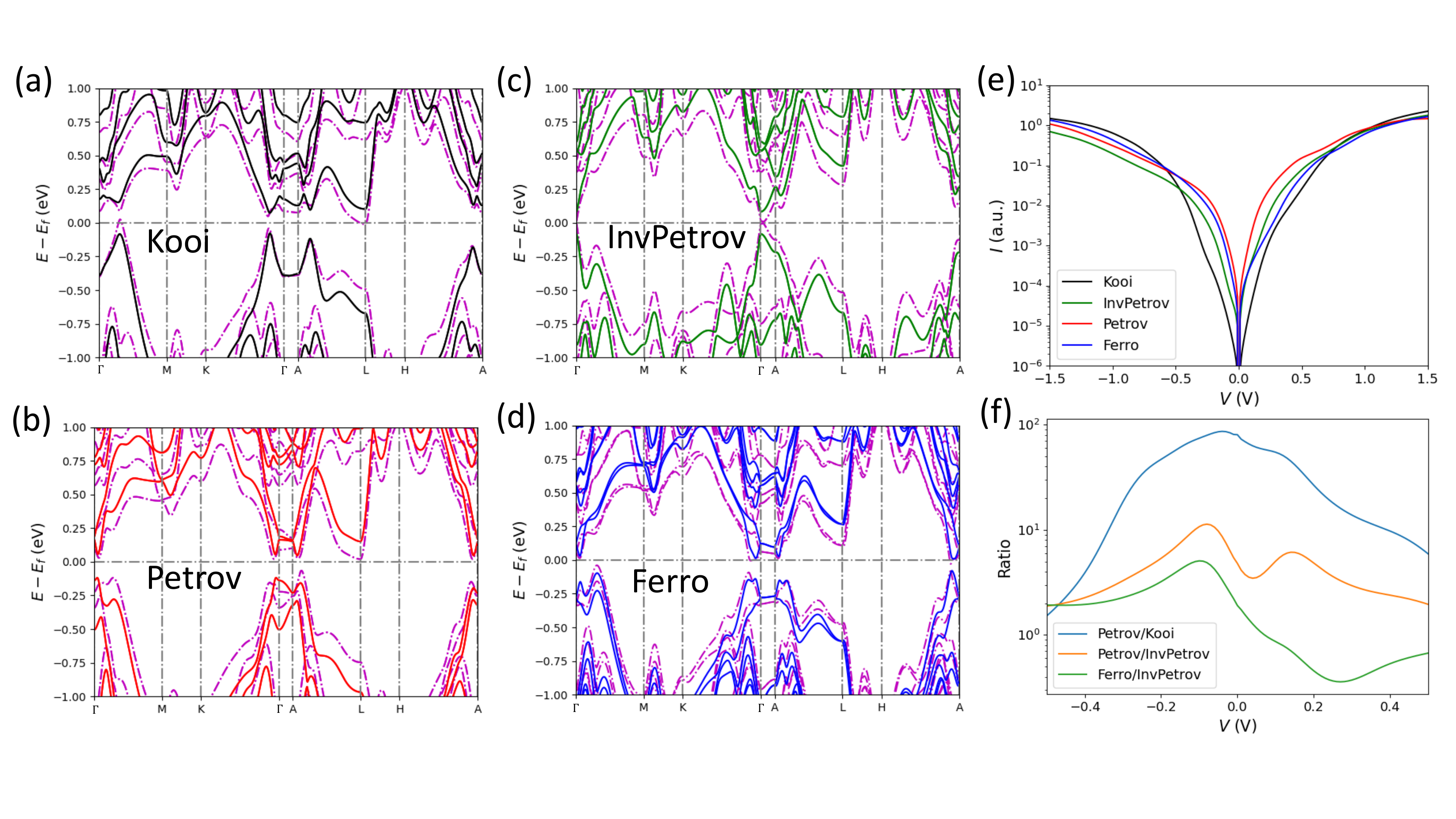}
 \caption{(a-d) Electronic structure obtained using HSE06 functional (solid lines) and PBE functional (magenta dashed lines) for (a) Kooi, (b) Petrov, (c) InvPetrov, and (d) Ferro phases respectively.
 (e) I-V curves of the four phases.
 (f) Ratio of the current flowing in two different phases as a function of the bias voltage.}
 \label{elec}
\end{figure*}

The atomic structures are relaxed using density functional theory (DFT) with PBE functional including VdW corrections (D3)\cite{Grimme_2010}, $8\times8\times4$ $k$-mesh, and $400$ eV cutoff energy.
Instability of Kooi structure with D3 correction (imaginary phonon) is observed, due to the larger volume of the unit cell, consistent with Ref.~\cite{Song_2018}. 
To avoid such instability, as reported in Ref.~\cite{Campi_2017}, we adopt D2 correction \cite{Grimme_2006} for Kooi structure.
The relaxed lattice parameters of hex unit cells ($a$ and $c$) are listed in Table~\ref{iPCM}. 
The relaxed structures are available in supplementary materials. 

Electronic structures are obtained using hybrid functional (HSE06)\cite{Heyd_2003} with $12\times12\times4$ $k$-mesh and $400$ eV cutoff energy.
Electron transport in the vertical direction (perpendicular to VdW gap) is computed using the non\dash{}equilibrium Green's function (NEGF) formalism based on maximally localized Wannier functions (MLWF) approach starting from HSE calculations.
The bias voltage is applied across the two electrodes, which are made of the same material.
MLWF was computed using Wannier90\cite{Pizzi_2020}, NEGF was performed using an in-house code\cite{Dragoni_2020}.

Phonon dispersions and anharmonicity are calculated using finite difference method with an atomic displacement of $0.01$ \AA ~and $0.03$ \AA, respectively i.e. $1472$ {\it{ab initio}} calculations after taking advantage of the structural symmetry for Kooi, Petrov, and Inverted\dash{}Petrov (InvPetrov); twice that amount for Ferro due to the breaking of bulk inversion symmetry. 
Phonon calculation adopts PBE functional with $400$ eV cut\dash{}off energy, $2\times2\times2$ supercell with $4\times4\times2$ $k$-mesh for phonon dispersion; $2\times2\times1$ super cell with $4\times4\times4$ $k$-mesh for anharmonicity calculation.
Thermal conductivity is calculated by solving the Boltzmann transport equation using the phono3py package \cite{phono3py} with interpolation on an $11\times11\times11$ $q$-mesh.
All {\it{ab initio}} calculations were performed using VASP code \cite{Kresse_1996,Kresse_1999} with spin\dash{}orbit coupling.

\begin{table*}[t]
  \caption{In-plane and out-of plane lattice parameter ($a$ and $c$) of hexagonal phases relaxed using PBE with VdW correction. 
  Electronic band gap obtained using HSE06 ($E_g^{\text{HSE06}}$) and PBE ($E_g^{\text{PBE}}$) functionals.
  In-plane and out-of-plane thermal conductivity ($\kappa_{\parallel}$ and $\kappa_{\perp}$) at $300$ K.
  Free energy with respect to Kooi structure at $300$ K.
  }
  \begin{tabular}{lccccccc}
  \hline\hline
  & a ($\AA$) & c ($\AA$) & $E_g^{\text{HSE06}}$ (meV) & $E_g^{\text{PBE}}$ (meV) & $\kappa_{\perp}$ (W m$^{-1}$K$^{-1}$) & $\kappa_{\parallel}$ (W m$^{-1}$K$^{-1}$) & $\Delta E$ (eV)\\ 
    \hline  
  Kooi      &4.213 & 17.190 & 166 & 11 & 1.526 & 0.450 & 0.0\\
  Petrov    &4.268 & 17.501 & 151 & 35 & 1.481 & 0.253 & 0.85\\
  InvPetrov &4.229 & 17.985 & 165 & 85 & 1.221 & 0.111 & 0.88\\
  Ferro	    &4.281 & 17.341 &  87 & 10 & 1.292 & 0.293 & 0.77\\
  \hline
  Exp. & 4.22$^a$ 4.27$^b$ &  17.24$^a$ 17.89$^b$ & - & - & \multicolumn{2}{c}{0.42$^c$ 0.47$^d$} & - \\
  \hline\hline
  $^a$ Ref.\cite{Matsunaga_2004} & $^b$ Ref. \cite{Park_2009}& $^c$ Ref. \cite{Lee_2013}& $^d$ Ref. \cite{Lyeo_2006}
  \end{tabular}
  \label{iPCM}
\end{table*}

Figure~\ref{elec}(a-d) shows the electronic structure of the four phases using PBE and HSE functionals.
The electronic gap calculated from PBE functional is about tens of meV, consistent with what was reported in Ref.~\cite{Tominaga_2014}. 
It is well known that PBE functional generally underestimates the electronic gap; HSE functional with a range-dependent fractional amount of Fock exchange is more reliable for the evaluation of the electronic structure \cite{Chen_2018}.
Indeed, HSE yields a larger bandgap with respect to PBE (Table \ref{iPCM}).
The four structures have a similar electronic gap, which is about $100$ meV.
The electronic gap has not been measured directly; an optical gap of $0.5$ eV has been reported, however, a smaller electronic gap of $0.3$ eV was considered to explain Hall measurements \cite{Lee_2005}.
Because of such a narrow electronic gap, we expect that these materials would behave like semi\dash{}metals with a relatively high electric conductivity at room temperature or beyond. 

\GeTeSbTe{} SL is stacked along $[0001]$ direction in a final device, with electrodes connected at the top and the bottom surfaces of the layer.
Therefore, electric transport in out\dash{}of\dash{}plane ($[0001]$) direction is modeled by NEGF method with the assumption of ballistic transport, which is suitable to evaluate the resistive contrast among different phases, thanks to the thin phase-change material (i.e. less than $100$~nm) and only a portion of the phase-change material participating in phase-transition, reducing further the transport channel length.
Fig.~\ref{elec}e shows the current as a function of voltage (I\dash{}V) characteristics at $300$~K computed for the four phases. 
The current flow is carried by conduction (valence) bands for positive (negative) bias voltages. 
We can observe strong similarities in the I\dash{}V characteristics of the four structures. The current ratio between different phases is reported in Fig.~\ref{elec}f in order to better highlight the differences. 
The maximum ratio of about $100$ is achieved between Petrov and Kooi structures.
However, the conductivity ratio between Petrov and InvPetrov, and between Ferro and InvPetrov phases are far below the one reported experimentally for the ratio between the SET and RESET states in PCM devices based on \GeTeSbTe SLs \cite{Mitrofanov_2019, Takaura_2014}, making it difficult to justify those results with a simplified hex\dash{}to\dash{}hex transition, as already mentioned in the introduction.
The low calculated resistive contrast is in good accordance with previous NEGF calculations with metal contacts, as reported in Ref.\cite{Nakamura_2017}.

\begin{figure*}
 \includegraphics[width=1.8\columnwidth]{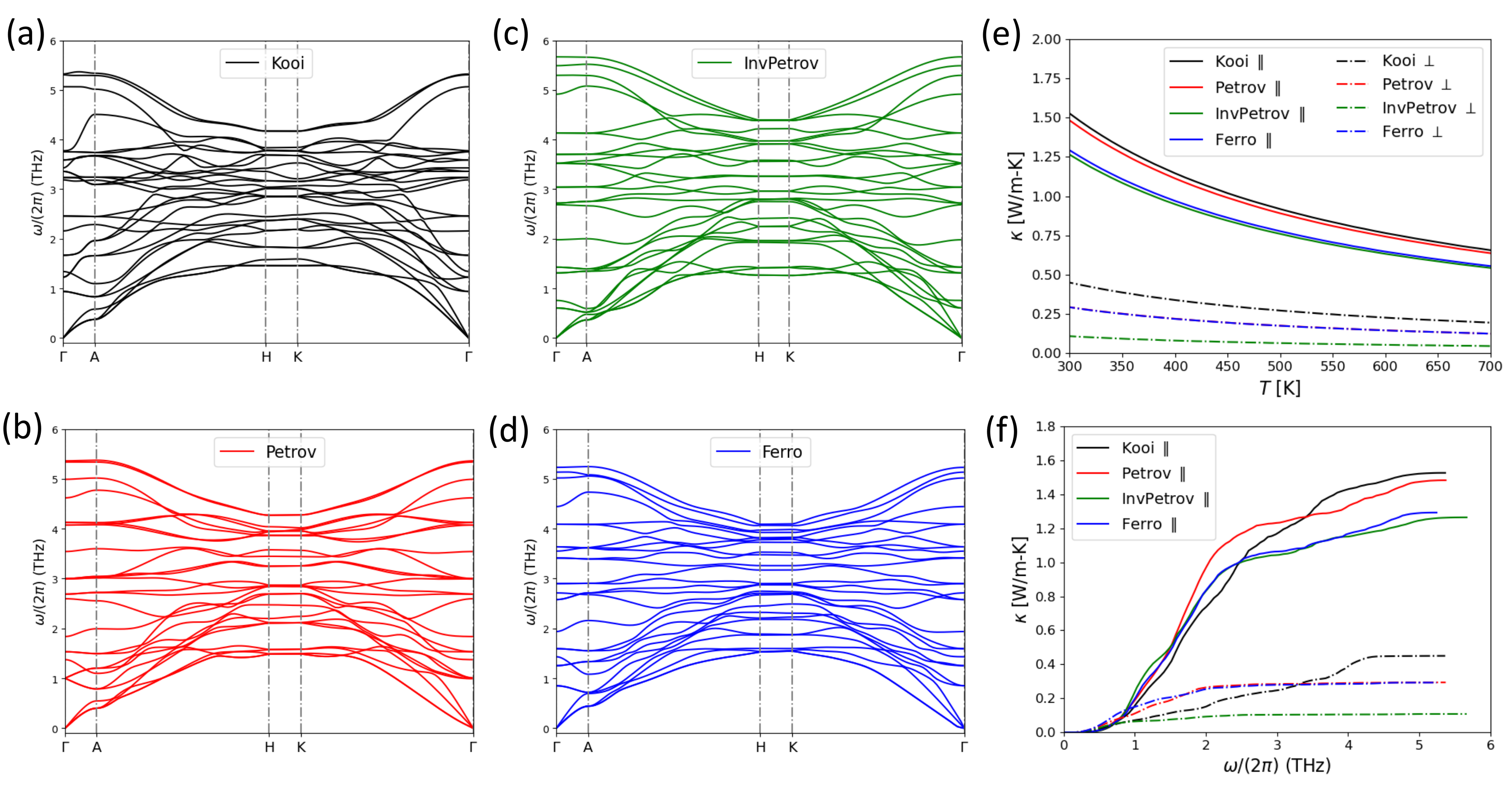}
 \caption{(a-d) Phonon dispersion for (a) Kooi, (b) Petrov, (c) InvPetrov, and (d) Ferro phases respectively.
 (e) Thermal conductivity of the four phases at different temperatures and (f) cumulative thermal conductivity, including phonons with frequency lower than $\omega$. In-plane and out-of-plane thermal conductivity are represented by solid lines and dashed lines respectively.
 }
 \label{thermal}
\end{figure*}

The phonon dispersions of the four structures are shown in Fig.~\ref{thermal}(a-d).
It is interesting to notice that phonon bands are less dispersive in the perpendicular direction ($\Gamma$-A) for \GeTeSbTe{} SLs than for Kooi structure.
This is likely because of the presence of additional VdW gaps and different bonds arrangement in SLs.
Indeed, in the unit cell of Petrov and Ferro phases, we find one Te\dash{}Te VdW gap and two Ge\dash{}Te long bonds, and InvPetrov phase presents two Te\dash{}Te VdW gaps and one Ge\dash{}Ge long bond. On the contrary, a single Te\dash{}Te VdW gap is present in Kooi structure.
These structural differences could have a huge impact on the out\dash{}of\dash{}plane thermal conductivity as shown in the following.

Thermal conductivities at different temperatures are shown in Fig.~\ref{thermal}e, and Table~\ref{iPCM}.
The in\dash{}plane thermal conductivity ($\kappa_{\parallel}$) is similar among the four structures and varies only from $1.25$ to $1.5$~W~m$^{-1}$K$^{-1}$ at $300$~K. 
The out\dash{}of\dash{}plane thermal conductivity ($\kappa_{\perp}$) varies of about $4$ times, from $0.11$ to $0.45$~W~m$^{-1}$K$^{-1}$ going from InvPetrov to Kooi structure, confirming the impact on thermal conductivity of the presence of different structural features between the four phases (i.e. VdW gaps and long bonds number). Our calculations confirm the anisotropic thermal conductivity of Kooi and Petrov phases reported by previous theoretical studies \cite{Mukhopadhyay_2016, Campi_2017}, and complete the picture adding InvPetrov and Ferro phases.
Previous experiments report a lattice thermal conductivity of about $0.42$\dash{}$0.47$~W~m$^{-1}$K$^{-1}$, which is close to the one we found for Kooi phase in the perpendicular direction \cite{Lyeo_2006, Lee_2013}.
The cumulative thermal conductivity shown in Fig.~\ref{thermal}f indicates that low energy phonons below $2$~THz dominate the out\dash{}of\dash{}plane thermal conductivity for \GeTeSbTe{} SLs. On the contrary, in Kooi structure the contribution from higher energy phonons at $4$~THz is visible.

Finally, by including phonons in free energy calculation, we find that Kooi phase is more stable than other three phases by about $0.8$~eV per unit cell at $300$~K (Table~\ref{iPCM}). 

Kooi structure is known to be the most stable among the hex GST structures, confirmed in our calculation of the free energy, and such stability can play a role during the growth of \GeTeSbTe{} SLs (i.e. during deposition) but also during the fabrication steps.
The intermixing of GeTe and SbTe layers, already observed in previous works, is likely to happen, in particular if SLs are submitted to high\dash{}temperature exposure during the fabrication process of the PCM device.
The hypothesis of a perfect crystal\dash{}to\dash{}crystal transition advanced to support the electrical results shown in literature, seems difficult to be sustained in the light of the poor conductivity contrast that we calculated between the four possible hex GST phases. As previously reported, our simulations corroborate the hypothesis that \GeTeSbTe{} SL devices could work like a standard PCM device, through a melt\dash{}quench process, but likely involving a smaller active volume in the PCM layer, which would reduce the energy consumption and increase the switching speed.
To better limit the size of the melted volume, it seems interesting to engineer the phase\dash{}change material stack: a) introducing highly conductive layers, to reduce the power consumption; b) finely controlling the material structure, adding structural features in order to enhance the thermal anisotropy of the layer (i.e. VdW gaps, long bonds, interfaces, etc.); c) carefully limiting/controlling the programming power to avoid the impact on the layer integrity.

In conclusion, we systematically investigated the electronic and thermal properties of hex GST phases, in particular comparing standard Kooi structure and \GeTeSbTe{} SLs (Petrov, InvPetrov, and Ferro). 
We demonstrated that all four phases have similar electronic properties, i.e. small bandgap (about $0.1$ eV) and semi\dash{}metallic behavior, leading to a low resistive contrast, which is not favorable to sustain a perfect crystal\dash{}to\dash{}crystal transition in SLs based PCM devices.
\GeTeSbTe{} SLs present a strong thermal conductivity anisotropy, correlated to the structure of the layer, nominally the presence of VdW gaps and long bonds (i.e. weak atomic interactions). 
Indeed, we show that all SLs structures present a lower out\dash{}of\dash{}plane thermal conductivity with respect to standard stable Kooi structure.
These results support the interest of the thermal engineering of the PCM cell starting from the fine\dash{}tuning of the structure of the phase\dash{}change layer.

{\textbf{Supplementary Material}}
Atomic coordinates in unit cells of Kooi, Petrov, InvPetrov, and Ferro phases are reported.

\begin{acknowledgments}
B. S. and J. L. thank the allocation of computational resource from GENCI–IDRIS (Grant 2020-A0090912036). 
\end{acknowledgments}

{\textbf{Data Availability}}
Data available on request from the authors.

\bibliography{GeTeSbTe}

\begin{thebibliography}{40}%
\makeatletter
\providecommand \@ifxundefined [1]{%
 \@ifx{#1\undefined}
}%
\providecommand \@ifnum [1]{%
 \ifnum #1\expandafter \@firstoftwo
 \else \expandafter \@secondoftwo
 \fi
}%
\providecommand \@ifx [1]{%
 \ifx #1\expandafter \@firstoftwo
 \else \expandafter \@secondoftwo
 \fi
}%
\providecommand \natexlab [1]{#1}%
\providecommand \enquote  [1]{``#1''}%
\providecommand \bibnamefont  [1]{#1}%
\providecommand \bibfnamefont [1]{#1}%
\providecommand \citenamefont [1]{#1}%
\providecommand \href@noop [0]{\@secondoftwo}%
\providecommand \href [0]{\begingroup \@sanitize@url \@href}%
\providecommand \@href[1]{\@@startlink{#1}\@@href}%
\providecommand \@@href[1]{\endgroup#1\@@endlink}%
\providecommand \@sanitize@url [0]{\catcode `\\12\catcode `\$12\catcode
  `\&12\catcode `\#12\catcode `\^12\catcode `\_12\catcode `\%12\relax}%
\providecommand \@@startlink[1]{}%
\providecommand \@@endlink[0]{}%
\providecommand \url  [0]{\begingroup\@sanitize@url \@url }%
\providecommand \@url [1]{\endgroup\@href {#1}{\urlprefix }}%
\providecommand \urlprefix  [0]{URL }%
\providecommand \Eprint [0]{\href }%
\providecommand \doibase [0]{http://dx.doi.org/}%
\providecommand \selectlanguage [0]{\@gobble}%
\providecommand \bibinfo  [0]{\@secondoftwo}%
\providecommand \bibfield  [0]{\@secondoftwo}%
\providecommand \translation [1]{[#1]}%
\providecommand \BibitemOpen [0]{}%
\providecommand \bibitemStop [0]{}%
\providecommand \bibitemNoStop [0]{.\EOS\space}%
\providecommand \EOS [0]{\spacefactor3000\relax}%
\providecommand \BibitemShut  [1]{\csname bibitem#1\endcsname}%
\let\auto@bib@innerbib\@empty
\bibitem [{\citenamefont {Cheng}\ \emph {et~al.}(2019)\citenamefont {Cheng},
  \citenamefont {Carta}, \citenamefont {Chien}, \citenamefont {Lung},\ and\
  \citenamefont {BrightSky}}]{Cheng_2019}%
  \BibitemOpen
  \bibfield  {author} {\bibinfo {author} {\bibfnamefont {H.-Y.}\ \bibnamefont
  {Cheng}}, \bibinfo {author} {\bibfnamefont {F.}~\bibnamefont {Carta}},
  \bibinfo {author} {\bibfnamefont {W.-C.}\ \bibnamefont {Chien}}, \bibinfo
  {author} {\bibfnamefont {H.-L.}\ \bibnamefont {Lung}}, \ and\ \bibinfo
  {author} {\bibfnamefont {M.~J.}\ \bibnamefont {BrightSky}},\ }\href {\doibase
  10.1088/1361-6463/ab39a0} {\bibfield  {journal} {\bibinfo  {journal} {Journal
  of Physics D: Applied Physics}\ }\textbf {\bibinfo {volume} {52}},\ \bibinfo
  {pages} {473002} (\bibinfo {year} {2019})}\BibitemShut {NoStop}%
\bibitem [{\citenamefont {Cappelletti}\ \emph {et~al.}(2020)\citenamefont
  {Cappelletti}, \citenamefont {Annunziata}, \citenamefont {Arnaud},
  \citenamefont {Disegni}, \citenamefont {Maurelli},\ and\ \citenamefont
  {Zuliani}}]{Cappelletti_2020}%
  \BibitemOpen
  \bibfield  {author} {\bibinfo {author} {\bibfnamefont {P.}~\bibnamefont
  {Cappelletti}}, \bibinfo {author} {\bibfnamefont {R.}~\bibnamefont
  {Annunziata}}, \bibinfo {author} {\bibfnamefont {F.}~\bibnamefont {Arnaud}},
  \bibinfo {author} {\bibfnamefont {F.}~\bibnamefont {Disegni}}, \bibinfo
  {author} {\bibfnamefont {A.}~\bibnamefont {Maurelli}}, \ and\ \bibinfo
  {author} {\bibfnamefont {P.}~\bibnamefont {Zuliani}},\ }\href {\doibase
  10.1088/1361-6463/ab71aa} {\bibfield  {journal} {\bibinfo  {journal} {Journal
  of Physics D: Applied Physics}\ }\textbf {\bibinfo {volume} {53}},\ \bibinfo
  {pages} {193002} (\bibinfo {year} {2020})}\BibitemShut {NoStop}%
\bibitem [{\citenamefont {Song}\ \emph {et~al.}(2020)\citenamefont {Song},
  \citenamefont {Cheng}, \citenamefont {Cai}, \citenamefont {Tang},
  \citenamefont {Song}, \citenamefont {Wang}, \citenamefont {Zhao},
  \citenamefont {Xin},\ and\ \citenamefont {Liu}}]{Song_2020}%
  \BibitemOpen
  \bibfield  {author} {\bibinfo {author} {\bibfnamefont {W.-X.}\ \bibnamefont
  {Song}}, \bibinfo {author} {\bibfnamefont {Y.}~\bibnamefont {Cheng}},
  \bibinfo {author} {\bibfnamefont {D.}~\bibnamefont {Cai}}, \bibinfo {author}
  {\bibfnamefont {Q.}~\bibnamefont {Tang}}, \bibinfo {author} {\bibfnamefont
  {Z.}~\bibnamefont {Song}}, \bibinfo {author} {\bibfnamefont {L.}~\bibnamefont
  {Wang}}, \bibinfo {author} {\bibfnamefont {J.}~\bibnamefont {Zhao}}, \bibinfo
  {author} {\bibfnamefont {T.}~\bibnamefont {Xin}}, \ and\ \bibinfo {author}
  {\bibfnamefont {Z.-P.}\ \bibnamefont {Liu}},\ }\href {\doibase
  10.1063/5.0011983} {\bibfield  {journal} {\bibinfo  {journal} {Journal of
  Applied Physics}\ }\textbf {\bibinfo {volume} {128}},\ \bibinfo {pages}
  {075101} (\bibinfo {year} {2020})}\BibitemShut {NoStop}%
\bibitem [{\citenamefont {Kusiak}\ \emph {et~al.}(2016)\citenamefont {Kusiak},
  \citenamefont {Battaglia}, \citenamefont {No{\'{e}}}, \citenamefont {Sousa},\
  and\ \citenamefont {Fillot}}]{Kusiak_2016}%
  \BibitemOpen
  \bibfield  {author} {\bibinfo {author} {\bibfnamefont {A.}~\bibnamefont
  {Kusiak}}, \bibinfo {author} {\bibfnamefont {J.-L.}\ \bibnamefont
  {Battaglia}}, \bibinfo {author} {\bibfnamefont {P.}~\bibnamefont
  {No{\'{e}}}}, \bibinfo {author} {\bibfnamefont {V.}~\bibnamefont {Sousa}}, \
  and\ \bibinfo {author} {\bibfnamefont {F.}~\bibnamefont {Fillot}},\ }\href
  {\doibase 10.1088/1742-6596/745/3/032104} {\bibfield  {journal} {\bibinfo
  {journal} {Journal of Physics: Conference Series}\ }\textbf {\bibinfo
  {volume} {745}},\ \bibinfo {pages} {032104} (\bibinfo {year}
  {2016})}\BibitemShut {NoStop}%
\bibitem [{\citenamefont {Simpson}\ \emph {et~al.}(2011)\citenamefont
  {Simpson}, \citenamefont {Fons}, \citenamefont {Kolobov}, \citenamefont
  {Fukaya}, \citenamefont {Krbal}, \citenamefont {Yagi},\ and\ \citenamefont
  {Tominaga}}]{Tominaga_2011}%
  \BibitemOpen
  \bibfield  {author} {\bibinfo {author} {\bibfnamefont {R.~E.}\ \bibnamefont
  {Simpson}}, \bibinfo {author} {\bibfnamefont {P.}~\bibnamefont {Fons}},
  \bibinfo {author} {\bibfnamefont {A.~V.}\ \bibnamefont {Kolobov}}, \bibinfo
  {author} {\bibfnamefont {T.}~\bibnamefont {Fukaya}}, \bibinfo {author}
  {\bibfnamefont {M.}~\bibnamefont {Krbal}}, \bibinfo {author} {\bibfnamefont
  {T.}~\bibnamefont {Yagi}}, \ and\ \bibinfo {author} {\bibfnamefont
  {J.}~\bibnamefont {Tominaga}},\ }\href {\doibase 10.1038/nnano.2011.96}
  {\bibfield  {journal} {\bibinfo  {journal} {Nature Nanotechnology}\ }\textbf
  {\bibinfo {volume} {6}},\ \bibinfo {pages} {501–505} (\bibinfo {year}
  {2011})}\BibitemShut {NoStop}%
\bibitem [{\citenamefont {Yu}\ and\ \citenamefont
  {Robertson}(2015)}]{Xiaoming_2015}%
  \BibitemOpen
  \bibfield  {author} {\bibinfo {author} {\bibfnamefont {X.}~\bibnamefont
  {Yu}}\ and\ \bibinfo {author} {\bibfnamefont {J.}~\bibnamefont {Robertson}},\
  }\href {\doibase https://doi.org/10.1038/srep12612} {\bibfield  {journal}
  {\bibinfo  {journal} {Scientific Reports}\ }\textbf {\bibinfo {volume} {5}},\
  \bibinfo {pages} {12612} (\bibinfo {year} {2015})}\BibitemShut {NoStop}%
\bibitem [{\citenamefont {Mitrofanov}\ \emph {et~al.}(2019)\citenamefont
  {Mitrofanov}, \citenamefont {Saito}, \citenamefont {Miyata}, \citenamefont
  {Fons}, \citenamefont {Kolobov},\ and\ \citenamefont
  {Tominaga}}]{Mitrofanov_2019}%
  \BibitemOpen
  \bibfield  {author} {\bibinfo {author} {\bibfnamefont {K.~V.}\ \bibnamefont
  {Mitrofanov}}, \bibinfo {author} {\bibfnamefont {Y.}~\bibnamefont {Saito}},
  \bibinfo {author} {\bibfnamefont {N.}~\bibnamefont {Miyata}}, \bibinfo
  {author} {\bibfnamefont {P.}~\bibnamefont {Fons}}, \bibinfo {author}
  {\bibfnamefont {A.~V.}\ \bibnamefont {Kolobov}}, \ and\ \bibinfo {author}
  {\bibfnamefont {J.}~\bibnamefont {Tominaga}},\ }\href {\doibase
  10.1002/pssr.201900105} {\bibfield  {journal} {\bibinfo  {journal} {physica
  status solidi (RRL) – Rapid Research Letters}\ }\textbf {\bibinfo {volume}
  {13}},\ \bibinfo {pages} {1900105} (\bibinfo {year} {2019})}\BibitemShut
  {NoStop}%
\bibitem [{\citenamefont {Kolobov}\ \emph {et~al.}(2017)\citenamefont
  {Kolobov}, \citenamefont {Fons}, \citenamefont {Saito},\ and\ \citenamefont
  {Tominaga}}]{Kolobov_2017}%
  \BibitemOpen
  \bibfield  {author} {\bibinfo {author} {\bibfnamefont {A.~V.}\ \bibnamefont
  {Kolobov}}, \bibinfo {author} {\bibfnamefont {P.}~\bibnamefont {Fons}},
  \bibinfo {author} {\bibfnamefont {Y.}~\bibnamefont {Saito}}, \ and\ \bibinfo
  {author} {\bibfnamefont {J.}~\bibnamefont {Tominaga}},\ }\href {\doibase
  10.1021/acsomega.7b00812} {\bibfield  {journal} {\bibinfo  {journal} {ACS
  Omega}\ }\textbf {\bibinfo {volume} {2}},\ \bibinfo {pages} {6223} (\bibinfo
  {year} {2017})}\BibitemShut {NoStop}%
\bibitem [{\citenamefont {Kowalczyk}\ \emph {et~al.}(2018)\citenamefont
  {Kowalczyk}, \citenamefont {Hippert}, \citenamefont {Bernier}, \citenamefont
  {Mocuta}, \citenamefont {Sabbione}, \citenamefont {Batista-Pessoa},\ and\
  \citenamefont {Noé}}]{Kowalczyk_2018}%
  \BibitemOpen
  \bibfield  {author} {\bibinfo {author} {\bibfnamefont {P.}~\bibnamefont
  {Kowalczyk}}, \bibinfo {author} {\bibfnamefont {F.}~\bibnamefont {Hippert}},
  \bibinfo {author} {\bibfnamefont {N.}~\bibnamefont {Bernier}}, \bibinfo
  {author} {\bibfnamefont {C.}~\bibnamefont {Mocuta}}, \bibinfo {author}
  {\bibfnamefont {C.}~\bibnamefont {Sabbione}}, \bibinfo {author}
  {\bibfnamefont {W.}~\bibnamefont {Batista-Pessoa}}, \ and\ \bibinfo {author}
  {\bibfnamefont {P.}~\bibnamefont {Noé}},\ }\href {\doibase
  10.1002/smll.201704514} {\bibfield  {journal} {\bibinfo  {journal} {Small}\
  }\textbf {\bibinfo {volume} {14}},\ \bibinfo {pages} {1704514} (\bibinfo
  {year} {2018})}\BibitemShut {NoStop}%
\bibitem [{\citenamefont {Chen}\ \emph
  {et~al.}(2018{\natexlab{a}})\citenamefont {Chen}, \citenamefont {Li},
  \citenamefont {Wang}, \citenamefont {Xie}, \citenamefont {Tian},
  \citenamefont {Zhang},\ and\ \citenamefont {Sun}}]{Chen2_2018}%
  \BibitemOpen
  \bibfield  {author} {\bibinfo {author} {\bibfnamefont {N.-K.}\ \bibnamefont
  {Chen}}, \bibinfo {author} {\bibfnamefont {X.-B.}\ \bibnamefont {Li}},
  \bibinfo {author} {\bibfnamefont {X.-P.}\ \bibnamefont {Wang}}, \bibinfo
  {author} {\bibfnamefont {S.-Y.}\ \bibnamefont {Xie}}, \bibinfo {author}
  {\bibfnamefont {W.~Q.}\ \bibnamefont {Tian}}, \bibinfo {author}
  {\bibfnamefont {S.}~\bibnamefont {Zhang}}, \ and\ \bibinfo {author}
  {\bibfnamefont {H.-B.}\ \bibnamefont {Sun}},\ }\href {\doibase
  10.1109/TNANO.2017.2779579} {\bibfield  {journal} {\bibinfo  {journal} {IEEE
  Transactions on Nanotechnology}\ }\textbf {\bibinfo {volume} {17}},\ \bibinfo
  {pages} {140–146} (\bibinfo {year} {2018}{\natexlab{a}})}\BibitemShut
  {NoStop}%
\bibitem [{\citenamefont {Saito}\ \emph {et~al.}(2019)\citenamefont {Saito},
  \citenamefont {Kolobov}, \citenamefont {Fons}, \citenamefont {Mitrofanov},
  \citenamefont {Makino}, \citenamefont {Tominaga},\ and\ \citenamefont
  {Robertson}}]{Saito_2019}%
  \BibitemOpen
  \bibfield  {author} {\bibinfo {author} {\bibfnamefont {Y.}~\bibnamefont
  {Saito}}, \bibinfo {author} {\bibfnamefont {A.~V.}\ \bibnamefont {Kolobov}},
  \bibinfo {author} {\bibfnamefont {P.}~\bibnamefont {Fons}}, \bibinfo {author}
  {\bibfnamefont {K.~V.}\ \bibnamefont {Mitrofanov}}, \bibinfo {author}
  {\bibfnamefont {K.}~\bibnamefont {Makino}}, \bibinfo {author} {\bibfnamefont
  {J.}~\bibnamefont {Tominaga}}, \ and\ \bibinfo {author} {\bibfnamefont
  {J.}~\bibnamefont {Robertson}},\ }\href {\doibase 10.1063/1.5088068}
  {\bibfield  {journal} {\bibinfo  {journal} {Applied Physics Letters}\
  }\textbf {\bibinfo {volume} {114}},\ \bibinfo {pages} {132102} (\bibinfo
  {year} {2019})}\BibitemShut {NoStop}%
\bibitem [{\citenamefont {Chong}\ \emph {et~al.}(2006)\citenamefont {Chong},
  \citenamefont {Shi}, \citenamefont {Zhao}, \citenamefont {Tan}, \citenamefont
  {Li}, \citenamefont {Lee}, \citenamefont {Miao}, \citenamefont {Du},\ and\
  \citenamefont {Tung}}]{Chong_2006}%
  \BibitemOpen
  \bibfield  {author} {\bibinfo {author} {\bibfnamefont {T.~C.}\ \bibnamefont
  {Chong}}, \bibinfo {author} {\bibfnamefont {L.~P.}\ \bibnamefont {Shi}},
  \bibinfo {author} {\bibfnamefont {R.}~\bibnamefont {Zhao}}, \bibinfo {author}
  {\bibfnamefont {P.~K.}\ \bibnamefont {Tan}}, \bibinfo {author} {\bibfnamefont
  {J.~M.}\ \bibnamefont {Li}}, \bibinfo {author} {\bibfnamefont {H.~K.}\
  \bibnamefont {Lee}}, \bibinfo {author} {\bibfnamefont {X.~S.}\ \bibnamefont
  {Miao}}, \bibinfo {author} {\bibfnamefont {A.~Y.}\ \bibnamefont {Du}}, \ and\
  \bibinfo {author} {\bibfnamefont {C.~H.}\ \bibnamefont {Tung}},\ }\href
  {\doibase 10.1063/1.2181191} {\bibfield  {journal} {\bibinfo  {journal}
  {Applied Physics Letters}\ }\textbf {\bibinfo {volume} {88}},\ \bibinfo
  {pages} {122114} (\bibinfo {year} {2006})}\BibitemShut {NoStop}%
\bibitem [{\citenamefont {Tong}\ \emph {et~al.}(2011)\citenamefont {Tong},
  \citenamefont {Miao}, \citenamefont {Cheng}, \citenamefont {Wang},
  \citenamefont {Zhang}, \citenamefont {Sun}, \citenamefont {Tong},\ and\
  \citenamefont {Wang}}]{Tong_2011}%
  \BibitemOpen
  \bibfield  {author} {\bibinfo {author} {\bibfnamefont {H.}~\bibnamefont
  {Tong}}, \bibinfo {author} {\bibfnamefont {X.~S.}\ \bibnamefont {Miao}},
  \bibinfo {author} {\bibfnamefont {X.~M.}\ \bibnamefont {Cheng}}, \bibinfo
  {author} {\bibfnamefont {H.}~\bibnamefont {Wang}}, \bibinfo {author}
  {\bibfnamefont {L.}~\bibnamefont {Zhang}}, \bibinfo {author} {\bibfnamefont
  {J.~J.}\ \bibnamefont {Sun}}, \bibinfo {author} {\bibfnamefont
  {F.}~\bibnamefont {Tong}}, \ and\ \bibinfo {author} {\bibfnamefont {J.~H.}\
  \bibnamefont {Wang}},\ }\href {\doibase 10.1063/1.3562610} {\bibfield
  {journal} {\bibinfo  {journal} {Applied Physics Letters}\ }\textbf {\bibinfo
  {volume} {98}},\ \bibinfo {pages} {101904} (\bibinfo {year}
  {2011})}\BibitemShut {NoStop}%
\bibitem [{\citenamefont {Long}, \citenamefont {Tong},\ and\ \citenamefont
  {Miao}(2012)}]{Long_2012}%
  \BibitemOpen
  \bibfield  {author} {\bibinfo {author} {\bibfnamefont {P.}~\bibnamefont
  {Long}}, \bibinfo {author} {\bibfnamefont {H.}~\bibnamefont {Tong}}, \ and\
  \bibinfo {author} {\bibfnamefont {X.}~\bibnamefont {Miao}},\ }\href {\doibase
  10.1143/apex.5.031201} {\bibfield  {journal} {\bibinfo  {journal} {Applied
  Physics Express}\ }\textbf {\bibinfo {volume} {5}},\ \bibinfo {pages}
  {031201} (\bibinfo {year} {2012})}\BibitemShut {NoStop}%
\bibitem [{\citenamefont {Campi}\ \emph {et~al.}(2017)\citenamefont {Campi},
  \citenamefont {Paulatto}, \citenamefont {Fugallo}, \citenamefont {Mauri},\
  and\ \citenamefont {Bernasconi}}]{Campi_2017}%
  \BibitemOpen
  \bibfield  {author} {\bibinfo {author} {\bibfnamefont {D.}~\bibnamefont
  {Campi}}, \bibinfo {author} {\bibfnamefont {L.}~\bibnamefont {Paulatto}},
  \bibinfo {author} {\bibfnamefont {G.}~\bibnamefont {Fugallo}}, \bibinfo
  {author} {\bibfnamefont {F.}~\bibnamefont {Mauri}}, \ and\ \bibinfo {author}
  {\bibfnamefont {M.}~\bibnamefont {Bernasconi}},\ }\href {\doibase
  10.1103/PhysRevB.95.024311} {\bibfield  {journal} {\bibinfo  {journal} {Phys.
  Rev. B}\ }\textbf {\bibinfo {volume} {95}},\ \bibinfo {pages} {024311}
  (\bibinfo {year} {2017})}\BibitemShut {NoStop}%
\bibitem [{\citenamefont {Navarro}\ \emph {et~al.}(2018)\citenamefont
  {Navarro}, \citenamefont {Bourgeois}, \citenamefont {Kluge}, \citenamefont
  {Serra}, \citenamefont {Verdy}, \citenamefont {Garrione}, \citenamefont
  {Cyrille}, \citenamefont {Bernier}, \citenamefont {Jannaud}, \citenamefont
  {Sabbione},\ and\ \citenamefont {et~al.}}]{Navarro_2018}%
  \BibitemOpen
  \bibfield  {author} {\bibinfo {author} {\bibfnamefont {G.}~\bibnamefont
  {Navarro}}, \bibinfo {author} {\bibfnamefont {G.}~\bibnamefont {Bourgeois}},
  \bibinfo {author} {\bibfnamefont {J.}~\bibnamefont {Kluge}}, \bibinfo
  {author} {\bibfnamefont {A.~L.}\ \bibnamefont {Serra}}, \bibinfo {author}
  {\bibfnamefont {A.}~\bibnamefont {Verdy}}, \bibinfo {author} {\bibfnamefont
  {J.}~\bibnamefont {Garrione}}, \bibinfo {author} {\bibfnamefont {M.-C.}\
  \bibnamefont {Cyrille}}, \bibinfo {author} {\bibfnamefont {N.}~\bibnamefont
  {Bernier}}, \bibinfo {author} {\bibfnamefont {A.}~\bibnamefont {Jannaud}},
  \bibinfo {author} {\bibfnamefont {C.}~\bibnamefont {Sabbione}}, \ and\
  \bibinfo {author} {\bibnamefont {et~al.}},\ }in\ \href {\doibase
  10.1109/IMW.2018.8388845} {\emph {\bibinfo {booktitle} {2018 IEEE
  International Memory Workshop (IMW)}}}\ (\bibinfo  {publisher} {IEEE},\
  \bibinfo {year} {2018})\ p.\ \bibinfo {pages} {1–4}\BibitemShut {NoStop}%
\bibitem [{\citenamefont {Boniardi}\ \emph {et~al.}(2019)\citenamefont
  {Boniardi}, \citenamefont {Boschker}, \citenamefont {Momand}, \citenamefont
  {Kooi}, \citenamefont {Redaelli},\ and\ \citenamefont
  {Calarco}}]{Boniardi_2019}%
  \BibitemOpen
  \bibfield  {author} {\bibinfo {author} {\bibfnamefont {M.}~\bibnamefont
  {Boniardi}}, \bibinfo {author} {\bibfnamefont {J.~E.}\ \bibnamefont
  {Boschker}}, \bibinfo {author} {\bibfnamefont {J.}~\bibnamefont {Momand}},
  \bibinfo {author} {\bibfnamefont {B.~J.}\ \bibnamefont {Kooi}}, \bibinfo
  {author} {\bibfnamefont {A.}~\bibnamefont {Redaelli}}, \ and\ \bibinfo
  {author} {\bibfnamefont {R.}~\bibnamefont {Calarco}},\ }\href {\doibase
  https://doi.org/10.1002/pssr.201800634} {\bibfield  {journal} {\bibinfo
  {journal} {physica status solidi (RRL) – Rapid Research Letters}\ }\textbf
  {\bibinfo {volume} {13}},\ \bibinfo {pages} {1800634} (\bibinfo {year}
  {2019})}\BibitemShut {NoStop}%
\bibitem [{\citenamefont {Térébénec}\ \emph {et~al.}(2021)\citenamefont
  {Térébénec}, \citenamefont {Castellani}, \citenamefont {Bernier},
  \citenamefont {Sever}, \citenamefont {Kowalczyk}, \citenamefont {Bernard},
  \citenamefont {Cyrille}, \citenamefont {Tran}, \citenamefont {Hippert},\ and\
  \citenamefont {Noé}}]{Noe_2021}%
  \BibitemOpen
  \bibfield  {author} {\bibinfo {author} {\bibfnamefont {D.}~\bibnamefont
  {Térébénec}}, \bibinfo {author} {\bibfnamefont {N.}~\bibnamefont
  {Castellani}}, \bibinfo {author} {\bibfnamefont {N.}~\bibnamefont {Bernier}},
  \bibinfo {author} {\bibfnamefont {V.}~\bibnamefont {Sever}}, \bibinfo
  {author} {\bibfnamefont {P.}~\bibnamefont {Kowalczyk}}, \bibinfo {author}
  {\bibfnamefont {M.}~\bibnamefont {Bernard}}, \bibinfo {author} {\bibfnamefont
  {M.-C.}\ \bibnamefont {Cyrille}}, \bibinfo {author} {\bibfnamefont {N.-P.}\
  \bibnamefont {Tran}}, \bibinfo {author} {\bibfnamefont {F.}~\bibnamefont
  {Hippert}}, \ and\ \bibinfo {author} {\bibfnamefont {P.}~\bibnamefont
  {Noé}},\ }\href {\doibase 10.1002/pssr.202000538} {\bibfield  {journal}
  {\bibinfo  {journal} {physica status solidi (RRL) – Rapid Research
  Letters}\ }\textbf {\bibinfo {volume} {15}},\ \bibinfo {pages} {2000538}
  (\bibinfo {year} {2021})}\BibitemShut {NoStop}%
\bibitem [{\citenamefont {Kwon}\ \emph {et~al.}(2021)\citenamefont {Kwon},
  \citenamefont {Khan}, \citenamefont {Perez}, \citenamefont {Asheghi},
  \citenamefont {Pop},\ and\ \citenamefont {Goodson}}]{Kwon_2021}%
  \BibitemOpen
  \bibfield  {author} {\bibinfo {author} {\bibfnamefont {H.}~\bibnamefont
  {Kwon}}, \bibinfo {author} {\bibfnamefont {A.~I.}\ \bibnamefont {Khan}},
  \bibinfo {author} {\bibfnamefont {C.}~\bibnamefont {Perez}}, \bibinfo
  {author} {\bibfnamefont {M.}~\bibnamefont {Asheghi}}, \bibinfo {author}
  {\bibfnamefont {E.}~\bibnamefont {Pop}}, \ and\ \bibinfo {author}
  {\bibfnamefont {K.~E.}\ \bibnamefont {Goodson}},\ }\href {\doibase
  10.1021/acs.nanolett.1c00947} {\bibfield  {journal} {\bibinfo  {journal}
  {Nano Letters}\ }\textbf {\bibinfo {volume} {21}},\ \bibinfo {pages}
  {5984–5990} (\bibinfo {year} {2021})}\BibitemShut {NoStop}%
\bibitem [{\citenamefont {Khan}\ \emph {et~al.}(2021)\citenamefont {Khan},
  \citenamefont {Daus}, \citenamefont {Islam}, \citenamefont {Neilson},
  \citenamefont {Lee}, \citenamefont {Wong},\ and\ \citenamefont
  {Pop}}]{Khan_2021}%
  \BibitemOpen
  \bibfield  {author} {\bibinfo {author} {\bibfnamefont {A.~I.}\ \bibnamefont
  {Khan}}, \bibinfo {author} {\bibfnamefont {A.}~\bibnamefont {Daus}}, \bibinfo
  {author} {\bibfnamefont {R.}~\bibnamefont {Islam}}, \bibinfo {author}
  {\bibfnamefont {K.~M.}\ \bibnamefont {Neilson}}, \bibinfo {author}
  {\bibfnamefont {H.~R.}\ \bibnamefont {Lee}}, \bibinfo {author} {\bibfnamefont
  {H.-S.~P.}\ \bibnamefont {Wong}}, \ and\ \bibinfo {author} {\bibfnamefont
  {E.}~\bibnamefont {Pop}},\ }\href {\doibase 10.1126/science.abj1261}
  {\bibfield  {journal} {\bibinfo  {journal} {Science}\ }\textbf {\bibinfo
  {volume} {373}},\ \bibinfo {pages} {1243–1247} (\bibinfo {year}
  {2021})}\BibitemShut {NoStop}%
\bibitem [{\citenamefont {Kooi}\ and\ \citenamefont
  {De~Hosson}(2002)}]{Kooi_2002}%
  \BibitemOpen
  \bibfield  {author} {\bibinfo {author} {\bibfnamefont {B.~J.}\ \bibnamefont
  {Kooi}}\ and\ \bibinfo {author} {\bibfnamefont {J.~T.~M.}\ \bibnamefont
  {De~Hosson}},\ }\href {\doibase 10.1063/1.1502915} {\bibfield  {journal}
  {\bibinfo  {journal} {Journal of Applied Physics}\ }\textbf {\bibinfo
  {volume} {92}},\ \bibinfo {pages} {3584–3590} (\bibinfo {year}
  {2002})}\BibitemShut {NoStop}%
\bibitem [{\citenamefont {Grimme}\ \emph {et~al.}(2010)\citenamefont {Grimme},
  \citenamefont {Antony}, \citenamefont {Ehrlich},\ and\ \citenamefont
  {Krieg}}]{Grimme_2010}%
  \BibitemOpen
  \bibfield  {author} {\bibinfo {author} {\bibfnamefont {S.}~\bibnamefont
  {Grimme}}, \bibinfo {author} {\bibfnamefont {J.}~\bibnamefont {Antony}},
  \bibinfo {author} {\bibfnamefont {S.}~\bibnamefont {Ehrlich}}, \ and\
  \bibinfo {author} {\bibfnamefont {H.}~\bibnamefont {Krieg}},\ }\href
  {\doibase 10.1063/1.3382344} {\bibfield  {journal} {\bibinfo  {journal} {The
  Journal of Chemical Physics}\ }\textbf {\bibinfo {volume} {132}},\ \bibinfo
  {pages} {154104} (\bibinfo {year} {2010})}\BibitemShut {NoStop}%
\bibitem [{\citenamefont {Song}, \citenamefont {Kim},\ and\ \citenamefont
  {Jhi}(2018)}]{Song_2018}%
  \BibitemOpen
  \bibfield  {author} {\bibinfo {author} {\bibfnamefont {Y.-S.}\ \bibnamefont
  {Song}}, \bibinfo {author} {\bibfnamefont {J.}~\bibnamefont {Kim}}, \ and\
  \bibinfo {author} {\bibfnamefont {S.-H.}\ \bibnamefont {Jhi}},\ }\href
  {\doibase 10.1103/PhysRevApplied.9.054044} {\bibfield  {journal} {\bibinfo
  {journal} {Physical Review Applied}\ }\textbf {\bibinfo {volume} {9}},\
  \bibinfo {pages} {054044} (\bibinfo {year} {2018})}\BibitemShut {NoStop}%
\bibitem [{\citenamefont {Grimme}(2006)}]{Grimme_2006}%
  \BibitemOpen
  \bibfield  {author} {\bibinfo {author} {\bibfnamefont {S.}~\bibnamefont
  {Grimme}},\ }\href {\doibase https://doi.org/10.1002/jcc.20495} {\bibfield
  {journal} {\bibinfo  {journal} {Journal of Computational Chemistry}\ }\textbf
  {\bibinfo {volume} {27}},\ \bibinfo {pages} {1787} (\bibinfo {year}
  {2006})}\BibitemShut {NoStop}%
\bibitem [{\citenamefont {Heyd}, \citenamefont {Scuseria},\ and\ \citenamefont
  {Ernzerhof}(2003)}]{Heyd_2003}%
  \BibitemOpen
  \bibfield  {author} {\bibinfo {author} {\bibfnamefont {J.}~\bibnamefont
  {Heyd}}, \bibinfo {author} {\bibfnamefont {G.~E.}\ \bibnamefont {Scuseria}},
  \ and\ \bibinfo {author} {\bibfnamefont {M.}~\bibnamefont {Ernzerhof}},\
  }\href {\doibase 10.1063/1.1564060} {\bibfield  {journal} {\bibinfo
  {journal} {The Journal of Chemical Physics}\ }\textbf {\bibinfo {volume}
  {118}},\ \bibinfo {pages} {8207} (\bibinfo {year} {2003})}\BibitemShut
  {NoStop}%
\bibitem [{\citenamefont {Pizzi}\ \emph {et~al.}(2020)\citenamefont {Pizzi},
  \citenamefont {Vitale}, \citenamefont {Arita}, \citenamefont {Blügel},
  \citenamefont {Freimuth}, \citenamefont {Géranton}, \citenamefont
  {Gibertini}, \citenamefont {Gresch}, \citenamefont {Johnson}, \citenamefont
  {Koretsune},\ and\ \citenamefont {et~al.}}]{Pizzi_2020}%
  \BibitemOpen
  \bibfield  {author} {\bibinfo {author} {\bibfnamefont {G.}~\bibnamefont
  {Pizzi}}, \bibinfo {author} {\bibfnamefont {V.}~\bibnamefont {Vitale}},
  \bibinfo {author} {\bibfnamefont {R.}~\bibnamefont {Arita}}, \bibinfo
  {author} {\bibfnamefont {S.}~\bibnamefont {Blügel}}, \bibinfo {author}
  {\bibfnamefont {F.}~\bibnamefont {Freimuth}}, \bibinfo {author}
  {\bibfnamefont {G.}~\bibnamefont {Géranton}}, \bibinfo {author}
  {\bibfnamefont {M.}~\bibnamefont {Gibertini}}, \bibinfo {author}
  {\bibfnamefont {D.}~\bibnamefont {Gresch}}, \bibinfo {author} {\bibfnamefont
  {C.}~\bibnamefont {Johnson}}, \bibinfo {author} {\bibfnamefont
  {T.}~\bibnamefont {Koretsune}}, \ and\ \bibinfo {author} {\bibnamefont
  {et~al.}},\ }\href {\doibase 10.1088/1361-648X/ab51ff} {\ \textbf {\bibinfo
  {volume} {32}},\ \bibinfo {pages} {165902} (\bibinfo {year}
  {2020})}\BibitemShut {NoStop}%
\bibitem [{\citenamefont {Dragoni}\ \emph {et~al.}(2020)\citenamefont
  {Dragoni}, \citenamefont {Sklénard}, \citenamefont {Olevano},\ and\
  \citenamefont {Triozon}}]{Dragoni_2020}%
  \BibitemOpen
  \bibfield  {author} {\bibinfo {author} {\bibfnamefont {A.}~\bibnamefont
  {Dragoni}}, \bibinfo {author} {\bibfnamefont {B.}~\bibnamefont {Sklénard}},
  \bibinfo {author} {\bibfnamefont {V.}~\bibnamefont {Olevano}}, \ and\
  \bibinfo {author} {\bibfnamefont {F.}~\bibnamefont {Triozon}},\ }\href
  {\doibase 10.1103/PhysRevB.101.075402} {\bibfield  {journal} {\bibinfo
  {journal} {Physical Review B}\ }\textbf {\bibinfo {volume} {101}},\ \bibinfo
  {pages} {075402} (\bibinfo {year} {2020})}\BibitemShut {NoStop}%
\bibitem [{\citenamefont {Togo}, \citenamefont {Chaput},\ and\ \citenamefont
  {Tanaka}(2015)}]{phono3py}%
  \BibitemOpen
  \bibfield  {author} {\bibinfo {author} {\bibfnamefont {A.}~\bibnamefont
  {Togo}}, \bibinfo {author} {\bibfnamefont {L.}~\bibnamefont {Chaput}}, \ and\
  \bibinfo {author} {\bibfnamefont {I.}~\bibnamefont {Tanaka}},\ }\href
  {\doibase 10.1103/PhysRevB.91.094306} {\bibfield  {journal} {\bibinfo
  {journal} {Phys. Rev. B}\ }\textbf {\bibinfo {volume} {91}},\ \bibinfo
  {pages} {094306} (\bibinfo {year} {2015})}\BibitemShut {NoStop}%
\bibitem [{\citenamefont {Kresse}\ and\ \citenamefont
  {Furthmüller}(1996)}]{Kresse_1996}%
  \BibitemOpen
  \bibfield  {author} {\bibinfo {author} {\bibfnamefont {G.}~\bibnamefont
  {Kresse}}\ and\ \bibinfo {author} {\bibfnamefont {J.}~\bibnamefont
  {Furthmüller}},\ }\href {\doibase 10.1103/PhysRevB.54.11169} {\bibfield
  {journal} {\bibinfo  {journal} {Physical Review B}\ }\textbf {\bibinfo
  {volume} {54}},\ \bibinfo {pages} {11169–11186} (\bibinfo {year}
  {1996})}\BibitemShut {NoStop}%
\bibitem [{\citenamefont {Kresse}\ and\ \citenamefont
  {Joubert}(1999)}]{Kresse_1999}%
  \BibitemOpen
  \bibfield  {author} {\bibinfo {author} {\bibfnamefont {G.}~\bibnamefont
  {Kresse}}\ and\ \bibinfo {author} {\bibfnamefont {D.}~\bibnamefont
  {Joubert}},\ }\href {\doibase 10.1103/PhysRevB.59.1758} {\bibfield  {journal}
  {\bibinfo  {journal} {Physical Review B}\ }\textbf {\bibinfo {volume} {59}},\
  \bibinfo {pages} {1758–1775} (\bibinfo {year} {1999})}\BibitemShut
  {NoStop}%
\bibitem [{\citenamefont {Matsunaga}, \citenamefont {Yamada},\ and\
  \citenamefont {Kubota}(2004)}]{Matsunaga_2004}%
  \BibitemOpen
  \bibfield  {author} {\bibinfo {author} {\bibfnamefont {T.}~\bibnamefont
  {Matsunaga}}, \bibinfo {author} {\bibfnamefont {N.}~\bibnamefont {Yamada}}, \
  and\ \bibinfo {author} {\bibfnamefont {Y.}~\bibnamefont {Kubota}},\ }\href
  {\doibase 10.1107/S0108768104022906} {\bibfield  {journal} {\bibinfo
  {journal} {Acta Crystallographica Section B: Structural Science}\ }\textbf
  {\bibinfo {volume} {60}},\ \bibinfo {pages} {685–691} (\bibinfo {year}
  {2004})}\BibitemShut {NoStop}%
\bibitem [{\citenamefont {Park}\ \emph {et~al.}(2009)\citenamefont {Park},
  \citenamefont {Eom}, \citenamefont {Lee}, \citenamefont {Da~Silva},
  \citenamefont {Kang}, \citenamefont {Lee},\ and\ \citenamefont
  {Khang}}]{Park_2009}%
  \BibitemOpen
  \bibfield  {author} {\bibinfo {author} {\bibfnamefont {J.-W.}\ \bibnamefont
  {Park}}, \bibinfo {author} {\bibfnamefont {S.~H.}\ \bibnamefont {Eom}},
  \bibinfo {author} {\bibfnamefont {H.}~\bibnamefont {Lee}}, \bibinfo {author}
  {\bibfnamefont {J.~L.~F.}\ \bibnamefont {Da~Silva}}, \bibinfo {author}
  {\bibfnamefont {Y.-S.}\ \bibnamefont {Kang}}, \bibinfo {author}
  {\bibfnamefont {T.-Y.}\ \bibnamefont {Lee}}, \ and\ \bibinfo {author}
  {\bibfnamefont {Y.~H.}\ \bibnamefont {Khang}},\ }\href {\doibase
  10.1103/PhysRevB.80.115209} {\bibfield  {journal} {\bibinfo  {journal}
  {Physical Review B}\ }\textbf {\bibinfo {volume} {80}},\ \bibinfo {pages}
  {115209} (\bibinfo {year} {2009})}\BibitemShut {NoStop}%
\bibitem [{\citenamefont {Lee}\ \emph {et~al.}(2013)\citenamefont {Lee},
  \citenamefont {Bozorg-Grayeli}, \citenamefont {Kim}, \citenamefont {Asheghi},
  \citenamefont {Philip~Wong},\ and\ \citenamefont {Goodson}}]{Lee_2013}%
  \BibitemOpen
  \bibfield  {author} {\bibinfo {author} {\bibfnamefont {J.}~\bibnamefont
  {Lee}}, \bibinfo {author} {\bibfnamefont {E.}~\bibnamefont {Bozorg-Grayeli}},
  \bibinfo {author} {\bibfnamefont {S.}~\bibnamefont {Kim}}, \bibinfo {author}
  {\bibfnamefont {M.}~\bibnamefont {Asheghi}}, \bibinfo {author} {\bibfnamefont
  {H.-S.}\ \bibnamefont {Philip~Wong}}, \ and\ \bibinfo {author} {\bibfnamefont
  {K.~E.}\ \bibnamefont {Goodson}},\ }\href {\doibase 10.1063/1.4807141}
  {\bibfield  {journal} {\bibinfo  {journal} {Applied Physics Letters}\
  }\textbf {\bibinfo {volume} {102}},\ \bibinfo {pages} {191911} (\bibinfo
  {year} {2013})}\BibitemShut {NoStop}%
\bibitem [{\citenamefont {Lyeo}\ \emph {et~al.}(2006)\citenamefont {Lyeo},
  \citenamefont {Cahill}, \citenamefont {Lee}, \citenamefont {Abelson},
  \citenamefont {Kwon}, \citenamefont {Kim}, \citenamefont {Bishop},\ and\
  \citenamefont {Cheong}}]{Lyeo_2006}%
  \BibitemOpen
  \bibfield  {author} {\bibinfo {author} {\bibfnamefont {H.-K.}\ \bibnamefont
  {Lyeo}}, \bibinfo {author} {\bibfnamefont {D.~G.}\ \bibnamefont {Cahill}},
  \bibinfo {author} {\bibfnamefont {B.-S.}\ \bibnamefont {Lee}}, \bibinfo
  {author} {\bibfnamefont {J.~R.}\ \bibnamefont {Abelson}}, \bibinfo {author}
  {\bibfnamefont {M.-H.}\ \bibnamefont {Kwon}}, \bibinfo {author}
  {\bibfnamefont {K.-B.}\ \bibnamefont {Kim}}, \bibinfo {author} {\bibfnamefont
  {S.~G.}\ \bibnamefont {Bishop}}, \ and\ \bibinfo {author} {\bibfnamefont
  {B.-k.}\ \bibnamefont {Cheong}},\ }\href {\doibase 10.1063/1.2359354}
  {\bibfield  {journal} {\bibinfo  {journal} {Applied Physics Letters}\
  }\textbf {\bibinfo {volume} {89}},\ \bibinfo {pages} {151904} (\bibinfo
  {year} {2006})}\BibitemShut {NoStop}%
\bibitem [{\citenamefont {Tominaga}\ \emph {et~al.}(2014)\citenamefont
  {Tominaga}, \citenamefont {Kolobov}, \citenamefont {Fons}, \citenamefont
  {Nakano},\ and\ \citenamefont {Murakami}}]{Tominaga_2014}%
  \BibitemOpen
  \bibfield  {author} {\bibinfo {author} {\bibfnamefont {J.}~\bibnamefont
  {Tominaga}}, \bibinfo {author} {\bibfnamefont {A.~V.}\ \bibnamefont
  {Kolobov}}, \bibinfo {author} {\bibfnamefont {P.}~\bibnamefont {Fons}},
  \bibinfo {author} {\bibfnamefont {T.}~\bibnamefont {Nakano}}, \ and\ \bibinfo
  {author} {\bibfnamefont {S.}~\bibnamefont {Murakami}},\ }\href {\doibase
  10.1002/admi.201300027} {\bibfield  {journal} {\bibinfo  {journal} {Advanced
  Materials Interfaces}\ }\textbf {\bibinfo {volume} {1}},\ \bibinfo {pages}
  {1300027} (\bibinfo {year} {2014})}\BibitemShut {NoStop}%
\bibitem [{\citenamefont {Chen}\ \emph
  {et~al.}(2018{\natexlab{b}})\citenamefont {Chen}, \citenamefont {Miceli},
  \citenamefont {Rignanese},\ and\ \citenamefont {Pasquarello}}]{Chen_2018}%
  \BibitemOpen
  \bibfield  {author} {\bibinfo {author} {\bibfnamefont {W.}~\bibnamefont
  {Chen}}, \bibinfo {author} {\bibfnamefont {G.}~\bibnamefont {Miceli}},
  \bibinfo {author} {\bibfnamefont {G.-M.}\ \bibnamefont {Rignanese}}, \ and\
  \bibinfo {author} {\bibfnamefont {A.}~\bibnamefont {Pasquarello}},\ }\href
  {\doibase 10.1103/PhysRevMaterials.2.073803} {\bibfield  {journal} {\bibinfo
  {journal} {Physical Review Materials}\ }\textbf {\bibinfo {volume} {2}},\
  \bibinfo {pages} {073803} (\bibinfo {year} {2018}{\natexlab{b}})}\BibitemShut
  {NoStop}%
\bibitem [{\citenamefont {Lee}\ \emph {et~al.}(2005)\citenamefont {Lee},
  \citenamefont {Abelson}, \citenamefont {Bishop}, \citenamefont {Kang},
  \citenamefont {Cheong},\ and\ \citenamefont {Kim}}]{Lee_2005}%
  \BibitemOpen
  \bibfield  {author} {\bibinfo {author} {\bibfnamefont {B.-S.}\ \bibnamefont
  {Lee}}, \bibinfo {author} {\bibfnamefont {J.~R.}\ \bibnamefont {Abelson}},
  \bibinfo {author} {\bibfnamefont {S.~G.}\ \bibnamefont {Bishop}}, \bibinfo
  {author} {\bibfnamefont {D.-H.}\ \bibnamefont {Kang}}, \bibinfo {author}
  {\bibfnamefont {B.-k.}\ \bibnamefont {Cheong}}, \ and\ \bibinfo {author}
  {\bibfnamefont {K.-B.}\ \bibnamefont {Kim}},\ }\href {\doibase
  10.1063/1.1884248} {\bibfield  {journal} {\bibinfo  {journal} {Journal of
  Applied Physics}\ }\textbf {\bibinfo {volume} {97}},\ \bibinfo {pages}
  {093509} (\bibinfo {year} {2005})}\BibitemShut {NoStop}%
\bibitem [{\citenamefont {Takaura}\ \emph {et~al.}(2014)\citenamefont
  {Takaura}, \citenamefont {Ohyanagi}, \citenamefont {Tai}, \citenamefont
  {Kinoshita}, \citenamefont {Akita}, \citenamefont {Morikawa}, \citenamefont
  {Shirakawa}, \citenamefont {Araidai}, \citenamefont {Shiraishi},
  \citenamefont {Saito},\ and\ \citenamefont {Tominaga}}]{Takaura_2014}%
  \BibitemOpen
  \bibfield  {author} {\bibinfo {author} {\bibfnamefont {N.}~\bibnamefont
  {Takaura}}, \bibinfo {author} {\bibfnamefont {T.}~\bibnamefont {Ohyanagi}},
  \bibinfo {author} {\bibfnamefont {M.}~\bibnamefont {Tai}}, \bibinfo {author}
  {\bibfnamefont {M.}~\bibnamefont {Kinoshita}}, \bibinfo {author}
  {\bibfnamefont {K.}~\bibnamefont {Akita}}, \bibinfo {author} {\bibfnamefont
  {T.}~\bibnamefont {Morikawa}}, \bibinfo {author} {\bibfnamefont
  {H.}~\bibnamefont {Shirakawa}}, \bibinfo {author} {\bibfnamefont
  {M.}~\bibnamefont {Araidai}}, \bibinfo {author} {\bibfnamefont
  {K.}~\bibnamefont {Shiraishi}}, \bibinfo {author} {\bibfnamefont
  {Y.}~\bibnamefont {Saito}}, \ and\ \bibinfo {author} {\bibfnamefont
  {J.}~\bibnamefont {Tominaga}},\ }in\ \href {\doibase
  10.1109/IEDM.2014.7047132} {\emph {\bibinfo {booktitle} {2014 IEEE
  International Electron Devices Meeting}}}\ (\bibinfo {year} {2014})\ pp.\
  \bibinfo {pages} {29.2.1--29.2.4}\BibitemShut {NoStop}%
\bibitem [{\citenamefont {Nakamura}\ \emph {et~al.}(2017)\citenamefont
  {Nakamura}, \citenamefont {Rungger}, \citenamefont {Sanvito}, \citenamefont
  {Inoue}, \citenamefont {Tominaga},\ and\ \citenamefont
  {Asai}}]{Nakamura_2017}%
  \BibitemOpen
  \bibfield  {author} {\bibinfo {author} {\bibfnamefont {H.}~\bibnamefont
  {Nakamura}}, \bibinfo {author} {\bibfnamefont {I.}~\bibnamefont {Rungger}},
  \bibinfo {author} {\bibfnamefont {S.}~\bibnamefont {Sanvito}}, \bibinfo
  {author} {\bibfnamefont {N.}~\bibnamefont {Inoue}}, \bibinfo {author}
  {\bibfnamefont {J.}~\bibnamefont {Tominaga}}, \ and\ \bibinfo {author}
  {\bibfnamefont {Y.}~\bibnamefont {Asai}},\ }\href {\doibase
  10.1039/C7NR03495D} {\bibfield  {journal} {\bibinfo  {journal} {Nanoscale}\
  }\textbf {\bibinfo {volume} {9}},\ \bibinfo {pages} {9386–9395} (\bibinfo
  {year} {2017})}\BibitemShut {NoStop}%
\bibitem [{\citenamefont {Mukhopadhyay}, \citenamefont {Lindsay},\ and\
  \citenamefont {Singh}(2016)}]{Mukhopadhyay_2016}%
  \BibitemOpen
  \bibfield  {author} {\bibinfo {author} {\bibfnamefont {S.}~\bibnamefont
  {Mukhopadhyay}}, \bibinfo {author} {\bibfnamefont {L.}~\bibnamefont
  {Lindsay}}, \ and\ \bibinfo {author} {\bibfnamefont {D.~J.}\ \bibnamefont
  {Singh}},\ }\href {\doibase 10.1038/srep37076} {\bibfield  {journal}
  {\bibinfo  {journal} {Scientific Reports}\ }\textbf {\bibinfo {volume} {6}},\
  \bibinfo {pages} {37076} (\bibinfo {year} {2016})}\BibitemShut {NoStop}%
\end{thebibliography}%

\end{document}